\documentclass[epj,nopacs]{svjour}
\pdfoutput=1
\usepackage{amsmath}
\usepackage{amsfonts}
\usepackage{amssymb}
\usepackage{color}
\usepackage{graphicx}
\usepackage{mathrsfs}
\usepackage{graphics}
\usepackage{ulem}
\usepackage{enumerate}
\begin{document}
%


\title{Split degenerate states and stable p+ip phases from holography
}
\author{Zhang-Yu Nie\inst{1,2}\thanks{e-mail:niezy@kmust.edu.cn} \and Qiyuan Pan\inst{3}\thanks{e-mail:panqiyuan@126.com} \and Hua-Bi Zeng\inst{4,5}\thanks{e-mail:zenghbi@gmail.com} \and Hui Zeng\inst{1,2}\thanks{e-mail:zenghui@kmust.edu.cn}
}
\mail {Zhang-Yu Nie}
\institute{Kunming University of Science and Technology, Kunming 650500, China \and 
Key Laboratory of Theoretical Physics, Institute of Theoretical Physics, Chinese Academy of Sciences, P.O.Box 2735, Beijing 100190, China \and
Department of Physics, Key Laboratory of Low Dimensional Quantum Structures and Quantum Control of Ministry of Education, and Synergetic Innovation Center for Quantum Effects and Applications, Hunan Normal University, Changsha, Hunan 410081, China\and
College of Physics Science and Technology, Yangzhou University, Jiangsu 225009, China\and
Department of Physics, National Central University, Chungli 32001, Taiwan}

\abstract{
In this paper, we investigate the p+$i$p superfluid phases in the complex vector field holographic p-wave model. We find that in the probe limit, the p+$i$p phase and the p-wave phase are equally stable, hence the p and $i$p orders can be mixed with an arbitrary ratio to form more general p+$\lambda i$p phases, which are also equally stable with the p-wave and p+$i$p phases. As a result, the system possesses a degenerate thermal state in the superfluid region. We further study the case with considering the back reaction on the metric, and find that the degenerate ground states will be separated into p-wave and p+$i$p phases, and the p-wave phase is more stable. Finally, due to the different critical temperature of the zeroth order phase transitions from p-wave and p+$i$p phases to the normal phase, there is a temperature region where the p+$i$p phase exists but the p-wave phase doesn't. In this region we find the stable holographic p+$i$p phase for the first time.
%
} 

\maketitle
%


\flushbottom

\section{\bf Introduction}

The AdS/CFT correspondence~\cite{Maldacena:1997re,Gubser:1998bc,Witten:1998qj} has been widely studied over the past years. As a strong-weak duality, it is believed to be a useful tool to study the strongly coupled systems in QCD~\cite{Sakai:2004cn,Karch:2006pv} as well as condensed matter physics~\cite{Zaanen:2015oix}. One of the most successful model of applying the AdS/CFT is the holographic superconductor model~\cite{Gubser:2008px,Hartnoll:2008vx,Herzog:2008he}, which opened an era for applying the
gauge/ gravity correspondence to solve the strongly coupled condensed matter problems. One can find more details in various AdS/CMT problems in the recently published book Ref.~\cite{Zaanen:2015oix}.

Since many high temperature superconductors are believed to be strongly coupled, the holographic tool has also been used to mimic the phase diagram of high Tc superconductors~\cite{Kiritsis:2015hoa,Chen:2016cym}. In these studies, an important issue is to deal with systems with multiple order parameters. One should study the competition and coexistence effect between the different orders to determine the final phase diagram of the system. Some previous studies already focused on this kind of holographic systems and produced many interesting results~\cite{Basu:2010fa,Cai:2013wma,Nie:2013sda,Huang:2011ac,DG,Krikun:2012yj,Donos:2012yu,Musso:2013ija,Nitti:2013xaa,Liu:2013yaa,Amado:2013lia,Amoretti:2013oia,Momeni:2013bca,Donos:2013woa}. A nice review including the development in this topic can be found in Ref.~\cite{Cai:2015cya}. 

Different from their s-wave cousin, the superconductors and superfluids with triplet pairing are more complicated. Such as in the superfluid phases of helium-3~\cite{Volhardt-Wolfle-1990}, the energy gap can either be isotropic in the B-phase, or be anisotropic in the A-phase. And the stabilization of the anisotropic A-phase is caused by some strong coupling effects. In order to theoretically study the laws of competition between the different orders with triplet pairing, we can consider realizing more triplet pairing phases holographically as a beginning. In another point of view, the realization of novel stable holographic phases is also important progress on the study of holographic duality.

One of the triplet pairing phases with great interest is the p+$i$p phase. In Ref.~\cite{Gubser:2008wv}, the author studied the p+$i$p phase in the SU(2) holographic p-wave model. But the calculation of quasi-normal modes tells us that the p+$i$p phase in that model is unstable, therefore, less people continued on studying the p+$i$p holographic phases, which always turn out to be unstable~\cite{Arias:2012py}. Since recently a new holographic p-wave model was proposed in Ref.~\cite{Cai:2013aca}, it is possible to setup the p+$i$p phase holographically in a new way, and in this way, the p+$i$p phase could be stable.

In this paper, we study the p+$i$p phase in the new holographic p-wave model, both in the probe limit and in the case considering the full backreaction. The rest of the paper is organized as follows. In Sec.~\ref{sect:setup} we give the set up of the new p-wave model~\cite{Cai:2013aca,Cai:2013pda,Li:2013rhw,Wu:2014bba,Pan:2015lit,Lai:2016yma,Nie:2014qma,Arias:2016nww} and introduce the ansatz for the p-wave and p+$i$p phases. In Sec.~\ref{sect:probe} we show the behavior of p-wave and p+$i$p condensates in probe limit, and show the existence of more general p+$\lambda i$p phases. We further study the case with considering the full back reaction on the metric in Sec.~\ref{sect:backer}. Finally, we conclude the main results in this paper and give some discussions in Sec.~\ref{sect:conclusion}.

\section{The holographic setup of p+$i$p phases in the new p-wave model}\label{sect:setup}
We start with the complex vector field holographic p-wave model~\cite{Cai:2013aca,Cai:2013pda,Li:2013rhw,Wu:2014bba,Pan:2015lit,Lai:2016yma,Nie:2014qma} and work in 4 dimensional bulk spacetime. The action of this model is
\begin{eqnarray}
S  &=&S_G+S_M,\\ \nonumber
S_G&=&\frac{1}{2 \kappa_g ^2}\int d^{4}x \sqrt{-g} \big(R-2\Lambda \big),\\ \nonumber
S_M&=&\frac{1}{q^2}\int d^{4}x \sqrt{-g}\Big(-\frac{1}{4}F_{\mu\nu}F^{\mu\nu}-\frac{1}{2} \rho^\dagger_{\mu\nu}\rho^{\mu\nu} 
-m_p^2 \rho^\dagger_\mu \rho^\mu \Big).
\end{eqnarray}
We write the action of the holographic model into the gravity section $S_G$ and the matter section $S_M$. In this action, $F_{\mu\nu}=\nabla_\mu A_\nu-\nabla_\nu A_\mu$ is the field strength for the U(1) gauge field $A_\mu$, $\rho_\mu$ is a complex vector field charged under the U(1) gauge field with charge $q$, and the superscript ``$^\dagger$" means complex conjugate. The field strength of $\rho_\mu$ is $\rho_{\mu\nu}=D_\mu \rho_\nu-D_\nu \rho_\mu$ with covariant derivative $D_\mu =\nabla_\mu -i A_\mu$. $m_p$ is the mass for the complex vector field and controls the conformal dimension of the p-wave order.

The equations of motion for the full system can be given by the equations for matter fields
\begin{eqnarray}
\nabla^\nu F_{\nu\mu} = i  (\rho^\nu\rho_{\nu\mu}^\dagger-\rho^{\nu\dagger}\rho_{\nu\mu}),\\
D^\nu \rho_{\nu\mu}  - m^2 \rho_\mu = 0,
\end{eqnarray}
and the Einstein equations
\begin{equation}
R_{\mu\nu} -\frac{1}{2}(R-2\Lambda) g_{\mu\nu} = b^2 \mathcal{T}_{\mu\nu},
\end{equation}
where $b= \kappa_g/q$ characterizes the strength of back reaction of the matter fields on the background geometry. $\mathcal{T}_{\mu\nu}$ is 
the stress-energy tensor of the matter sector
\begin{eqnarray}
\mathcal{T}_{\mu\nu} = &(-\frac{1}{4}F_{\mu\nu}^a F^{a\mu\nu} -\frac{1}{2} \rho^\dagger_{\mu\nu|}\rho^{\mu\nu}-m_p^2 \rho^\dagger_\mu \Psi^\mu) g_{\mu\nu} +F_{\mu\lambda} F_\nu^\lambda  \nonumber\\
& + \rho^\dagger_{\mu\lambda}\rho_\nu^\lambda+\rho^\dagger_{\nu\lambda}\rho_\mu^\lambda+m_p^2 (\rho^\dagger_\mu \rho_\nu+ \rho^\dagger_\nu \rho_\mu).
\end{eqnarray}

In order to realize the p-wave and p+$i$p solutions, we take the ansatz for matter fields as
\begin{eqnarray}
A_t=\phi(r),~\rho_x=\Psi_x(r),~\rho_y=i \Psi_y(r).
\end{eqnarray}
In general, the $\Psi_x(r)$ and $\Psi_y(r)$ fields can be independent, therefore the metric form will also be anisotropic in general. A consistent metric form~\cite{Nie:2014qma,Ammon:2009xh} can be given as
\begin{eqnarray}\label{fullmetric}
&ds^2=&-N(r) \sigma (r)^2dt^2+\frac{1}{N(r)}dr^2 \\ \nonumber
&&+\frac{r^2}{L^2}(\frac{1}{f(r)^2}dx^{2}+f(r)^2 dy^2),
\end{eqnarray}
with
\begin{equation}
N(r)=\frac{r^2}{L^2}(1-\frac{2 M(r)}{r^3}),
\end{equation}
where $L$ is the AdS radius and is related to the cosmological constant by the relation $\Lambda=-3/L^2$.

Under the above matter and metric ansatz, the equations of motion can be written as
\begin{eqnarray}\label{EoMs1}
M'(r)&=&
\frac{b^2 L^4 \phi (r)^2}{2 N(r) \sigma (r)^2}  \Big(f(r)^2 \Psi_x(r)^2+\frac{\Psi_y(r)^2}{f(r)^2} \Big) \\&& \nonumber
+\frac{1}{2} b^2 L^4  N(r)   \Big(f(r)^2\Psi_x'(r)^2 +\frac{\Psi_y'(r)^2}{f(r)^2} \Big) \\&& \nonumber
+\frac{1}{2}b^2 L^4 m_p^2  (f(r)^2 \Psi_x(r)^2+\frac{\Psi_y(r)^2}{f(r)^2})  \\&&
+\frac{b^2 L^2 r^2 \phi '(r)^2}{4 \sigma (r)^2} +\frac{L^2 r^2 N(r) f'(r)^2}{2 f(r)^2}, \nonumber
\\
\label{EoMs2} \sigma '(r)&=&\frac{b^2 L^2 \phi (r)^2}{r N(r)^2 \sigma (r)}\Big(f(r)^2 \Psi_x(r)^2+\frac{\Psi_y(r)^2}{f(r)^2} \Big) \\&& \nonumber
+\frac{b^2 L^2 \sigma (r) }{r}\Big(f(r)^2\Psi_x'(r)^2 +\frac{\Psi_y'(r)^2}{f(r)^2} \Big)  \nonumber\\&&
+\frac{r \sigma (r) f'(r)^2}{f(r)^2}, \nonumber
\\
\label{EoMs3} f''(r)&=&-\frac{b^2 L^2 f(r) \phi (r)^2}{r^2 N(r)^2 \sigma (r)^2} \Big(f(r)^2 \Psi_x(r)^2-\frac{\Psi_y(r)^2}{f(r)^2} \Big) \\&& \nonumber
+\frac{b^2 L^2 f(r)}{r^2}\Big(f(r)^2\Psi_x'(r)^2 -\frac{\Psi_y'(r)^2}{f(r)^2} \Big)\nonumber\\&&
+\frac{b^2 L^2 m_p^2 f(r)}{r^2 N(r)}\Big(f(r)^2 \Psi_x(r)^2-\frac{\Psi_y(r)^2}{f(r)^2} \Big)\nonumber\\&&
+\frac{f'(r)^2}{f(r)}
-\frac{f'(r) N'(r)}{N(r)}-\frac{f'(r) \sigma '(r)}{\sigma (r)}-\frac{2 f'(r)}{r},  \nonumber\\
\label{EoMs4} \phi ''(r)&=&\Big(\frac{\sigma '(r)}{\sigma (r)}-\frac{2 }{r} \Big)\phi '(r) \\&& \nonumber
+\frac{2 L^2}{r^2 N(r)}\Big(f(r)^2 \Psi_x(r)^2+\frac{\Psi_y(r)^2}{f(r)^2} \Big)  \phi (r),\\
\label{EoMs5} \Psi_x''(r)&=&-\Big( \frac{N'(r)}{N(r)}+\frac{\sigma'(r)}{\sigma(r)}+\frac{2 f'(r)}{f(r)} \Big) \Psi_x'(r) \\&& \nonumber
-\Big( \frac{\phi(r)^2}{N(r)^2 \sigma(r)^2}-\frac{m_p^2}{N(r)} \Big) \Psi_x(r), \\
\label{EoMs6} \Psi_y''(r)&=&-\Big( \frac{N'(r)}{N(r)}+\frac{\sigma'(r)}{\sigma(r)}-\frac{2 f'(r)}{f(r)} \Big) \Psi_y'(r) \\&& \nonumber
-\Big( \frac{\phi(r)^2}{N(r)^2 \sigma(r)^2}-\frac{m_p^2}{N(r)} \Big) \Psi_y(r).
\end{eqnarray}

Note that the above equations admits the following four sets of scaling symmetries
\begin{eqnarray}
(1)& \phi \rightarrow \lambda^2 \phi~,~\Psi_x \rightarrow \lambda^2 \Psi_x~,~\Psi_y \rightarrow \lambda^2 \Psi_y~,~m_p\rightarrow \lambda m_p~,\nonumber\\&
N \rightarrow \lambda^2 N~,~L\rightarrow \lambda^{-1} L~,~b \rightarrow \lambda^{-1} b~;
\\(2)& \phi \rightarrow \lambda \phi~,~\Psi_x \rightarrow \lambda \Psi_x~,~\Psi_y \rightarrow \lambda \Psi_y~,~N \rightarrow \lambda^2 N~,\nonumber\\&
~M \rightarrow \lambda^3 M~,~r \rightarrow \lambda r~;
\\(3)& \phi \rightarrow \lambda \phi~,~\sigma \rightarrow \lambda \sigma~; \label{scaling3}
\\(4)& \Psi_x \rightarrow \lambda \Psi_x~,~\Psi_y \rightarrow \lambda^{-1} \Psi_y~,~f \rightarrow \lambda^{-1} f~. \label{scaling4}
\end{eqnarray}
These scaling symmetries will be used to simplify the numerical work. We will use the first scaling symmetry to set $L=1$ and the second one to set $r_h=1$. After we get the numerical solutions, we can use these scaling symmetries again to recover $L$ and $r_h$ to any value. The last two scaling symmetries will be used to rescale any solution to be asymptotically AdS, which means $\lim\limits_{r\to\infty} \sigma(r) \rightarrow1$ and $\lim\limits_{r\to\infty}f(r)\rightarrow 1$. 

In order to solve the equations of motion numerically, we need to specify the boundary conditions both on the horizon $r=r_h$ and on the boundary $r=\infty$. Without loss of generality, we set $L=1$ in the rest of the paper. Then near the horizon the functions can be expanded as

\begin{eqnarray}
M(r) &=& \frac{r_h^3}{2}+M_{h1} (r-r_h) + ...~,\\
\sigma(r) &=& \sigma_{h0}+\sigma_{h1}(r-r_h)+...~,\\
f(r) &=& f_{h0} + f_{h1}(r-r_h)+...~,\\
\phi(r) &=& \phi_{h1}(r-r_h)+\phi_{h2}(r-r_h)^2+...~,\\
\Psi_x(r) &=& \Psi_{xh0}+\Psi_{xh1}(r-r_h)+...~,\\
\Psi_y(r) &=& \Psi_{yh0}+\Psi_{yh1}(r-r_h)+...~.
\end{eqnarray}

One can check that only the coefficients $\{\sigma_{h0},f_{h0},\phi_{h1},\Psi_{xh0}, \Psi_{yh0} \}$ are independent. We also need to expand the functions near the  AdS boundary
\begin{eqnarray}
M(r)=& M_{b0}+ \frac{M_{b1}}{r} + ...~,~
&\sigma(r)=\sigma_{b0}+ \frac{\sigma_{b3}}{r^3}+...~, \nonumber\\
f(r)=& f_{b0} +  \frac{f_{b3}}{r^3}+...~,~
&\phi(r)=\mu - \frac{\rho }{r}+...~, \\ \nonumber
\Psi_x(r)=& \frac{\Psi_{x-}}{r^{1-\Delta}}+  \frac{\Psi_{x+}}{r^{\Delta}}+...~,~
&\Psi_y(r)= \frac{\Psi_{y-}}{r^{1-\Delta}}+  \frac{\Psi_{y+}}{r^{\Delta}}+...~,
\end{eqnarray}
where
\begin{equation}
\Delta=(1+\sqrt{1+4m_p^2 L^2})/2
\end{equation}
is the scaling dimension of the vector order.

In order to make the boundary geometry to be asymptotically AdS, we should have $\sigma_{b0}=f_{b0}=1$. These two conditions could be easily satisfied by a scale transformation from any known solution by using the last two scaling symmetries (\ref{scaling3},\ref{scaling4}). In the asymptotically  AdS spacetime, we can use the AdS/CFT correspondence to get information of the dual field theory. The AdS/CFT dictionary tells us that $\mu$ and $\rho$ are related to the chemical potential and charge density respectively, while $\Psi_{xb0},\Psi_{yb0}$ are related to the sources and $\Psi_{xb1},\Psi_{yb1}$ are related to the expectation values of the dual vector operators. Since we focus on the solutions with no source term, we further set $\Psi_{x-}=\Psi_{y-}=0$ as additional constraints.

With these four constraints on boundary, we only have one free parameter left for the solutions. In other words, we can get one parameter solutions that mimic the p-wave or p+$i$p phase transition holographically.

Before we go to the general case with considering the full interactions between matter fields and the metric, we show some interesting results in the probe limit where we can ignore the influence of matter fields on the metric. This limit can be taken consistently if we set $b\rightarrow 0$ or equivalently $q\rightarrow \infty$. We show the results in the next section and come back to the general case with finite $b$ and $q$ in Sec.~\ref{sect:backer}.

\section{Degenerate states in the probe limit}\label{sect:probe}
In this section, we take the probe limit with $b\rightarrow 0$. In this limit, the matter fields will not affect the metric fields, and the background metric (\ref{fullmetric}) becomes the analytical form of black brane solution
\begin{equation}
N(r)=\frac{r^2}{L^2}\Big(1-\frac{r_h^3}{r^3}~\Big),~ ~ ~\sigma(r)=f(r)=1.
\end{equation}
And the temperature of this solution is
\begin{equation}\label{Temperature}
T=\frac{3}{4 \pi L^2} r_h.
\end{equation}

We only need to solve the following equations of motion for matter fields in the AdS black brane background

\begin{eqnarray}
\phi''+\frac{2}{r}\phi' -2 L^2 \frac{\Psi_x^2+\Psi_y^2}{r^2 N}\phi&=&0, \label{eqphi}\\
\Psi_x''+\frac{N'}{N}\Psi_x'+\frac{\phi^2-m_p^2 f}{N^2}\Psi_x&=&0, \label{eqpsix}\\
\Psi_y''+\frac{N'}{N}\Psi_y'+\frac{\phi^2-m_p^2 f}{N^2}\Psi_y&=&0. \label{eqpsiy}
\end{eqnarray}
We can see that in these equations, $\Psi_x$ and $\Psi_y$ are not directly coupled in their equations of motion, and they both coupled to the same U(1) electromagnetic field. Thus we can consistently set $\Psi_x=0$ or $\Psi_y=0$, and in either case, we get a holographic p-wave superfluid phase. In order to get a p+$i$p phase, we should solve the three equations of motion to get a solution with $\Psi_x=\Psi_y$. The boundary conditions of these three fields are the same as in the general case given in the previous section.

\subsection{The p-wave solutions in the probe limit} \label{subsect:p-wave}
If we turn off $\Psi_y$, one can solve the remaining two equations of motion numerically to get p-wave solutions which are dual to the p-wave superfluid phases. These different p-wave solutions can be labeled by the values of $\Delta$ and $T$. For later convenience, we define these solutions as
\begin{equation}\label{pwave}
\Psi_x=\Psi_{\Delta,T}~,\quad \Psi_y=0~,\quad \phi=\phi_{\Delta,T}.
\end{equation}
$\Psi_{\Delta,T}$ and $\phi_{\Delta,T}$ satisfy the equations of motion \eqref{eqphi},\eqref{eqpsix} with $\Psi_y=0$ at the giving value of $\{\Delta,T\}$. From these solutions, we can study the behavior of the p-wave condensate to see the properties of p-wave phase transitions. It is easy to check that the condensate behaviors of the p-wave superfluid phases in the probe limit at different values of $\Delta$ are all similar to those in the SU(2) holographic p-wave model. The main difference is that we can tune the dimension of the p-wave order in this model, and different values of $\Delta$ will make the critical temperature of the p-wave superfluid phase transitions also different.

\subsection{Constructing p+$i$p and the general p+$\lambda i$p solutions from the p-wave solution} \label{subsect:p+ip}
We can also solve the three equations of motion with $\Psi_x=\Psi_y$ to get holographic dual of p+$i$p solutions on the boundary field theory. Here wee need not to do the numerics again, because these solutions can be easily constructed from the p-wave solutions in hand as
\begin{equation}\label{p+ip}
\Psi_x=\Psi_y=\frac{1}{\sqrt{2}}\Psi_{\Delta,T}~,\quad \phi=\phi_{\Delta,T}.
\end{equation}
One can check these solutions by substituting them into the three equations of motion, and get the following three equations
\begin{eqnarray}
\phi_{\Delta,T}''+\frac{2}{r}\phi_{\Delta,T}' -2 L^2 \frac{\Psi_{\Delta,T}}{r^2 N}\phi_{\Delta,T}&=&0, \label{checkEoM1}\\ 
\Psi_{\Delta,T}''+\frac{N'}{N}\Psi_{\Delta,T}'+\frac{\phi_{\Delta,T}^2-m_p^2 f}{N^2}\Psi_{\Delta,T}&=&0,\label{checkEoM2}\\ 
\Psi_{\Delta,T}''+\frac{N'}{N}\Psi_{\Delta,T}'+\frac{\phi_{\Delta,T}^2-m_p^2 f}{N^2}\Psi_{\Delta,T}&=&0.\label{checkEoM3}
\end{eqnarray}
These equations are always satisfied because they are just the two equations of motion \eqref{eqphi},\eqref{eqpsix} satisfied by the p-wave solutions $\{\Psi_{\Delta,T}~,~\phi_{\Delta,T}\}$.

We can go further to get solutions with $\Psi_x\neq\Psi_y$.  In these cases, from the boundary field theory point of view, the expectation value of the p-wave order is different to that of the $i$p order. We call these solutions as p+$\lambda i$p solutions and these solutions can be constructed as
\begin{equation}\label{p+lip}
\Psi_x=cos\theta~\Psi_{\Delta,T}~,\quad \Psi_y=sin\theta~\Psi_{\Delta,T}~,\quad \phi=\phi_{\Delta,T}~,
\end{equation}
where $\theta$ is an arbitrary valued constant angle. We can again check these solutions by substituting them into the equations of motion and get Eqs.~\eqref{checkEoM1},\eqref{checkEoM2},\eqref{checkEoM3}. From these solutions we can see that the p-wave solutions and p+$i$p solutions are special cases of the p+$\lambda i$ p solutions with $\theta=0$ and $\theta=\pi/4$ respectively.

\subsection{Free Energy in probe limit and the stability of p+$\lambda i$p phases} \label{subsect:freeE}
We have constructed the p+$\lambda i$p solutions in the previous section, and with $\theta=\pi/4$ we can get solutions dual to the p+$i$p superfluid phases. It is important to check wether these new solutions are stable or not. We can examine this stability by calculating the free energy of the phases with different $\lambda$.

In this paper, we work in the grand canonical potential, and the Gibbs free energy of the holographic system can be evaluated by the on-shell Euclidean action of the bulk system. Because we work in the probe limit, at a given temperature, the difference of the Gibbs free energy of the different phases only comes from the contribution of the matter part of the Euclidean action $S_{ME}$. The matter contribution to the Gibbs free energy of the holographic system is

\begin{eqnarray}
\Omega_m =T S_{ME}=\frac{V_2}{q^2}\Big(-\frac{1}{2}\mu \rho + \int_{r_h}^\infty \frac{\Psi_x^2+\Psi_y^2}{N}\phi^2 dr \Big).
\end{eqnarray}
We can substitute the p+$\lambda i$p solutions \eqref{p+lip} into the above formula to calculate the free energy of the different phases including the p-wave and p+$i$p one. It is easy to see that, at fixed value of $\{\Delta,~T\}$, the matter contribution of free energy for the p+$\lambda i$p phases equals to
\begin{eqnarray}
\Omega_m^{p+\lambda i p} = \frac{V_2}{q^2}\Big(-\frac{1}{2}\mu_{\Delta,T} \rho_{\Delta,T} + \int_{r_h}^\infty \frac{\Psi_{\Delta,T}}{N}\phi_{\Delta,T}^2 dr \Big),
\end{eqnarray}
where $\mu_{\Delta,T}$ and $\rho_{\Delta,T}$ are read from the boundary behavior of the function $\phi_{\Delta,T}$.

\begin{figure}\center
\includegraphics[width=8cm] {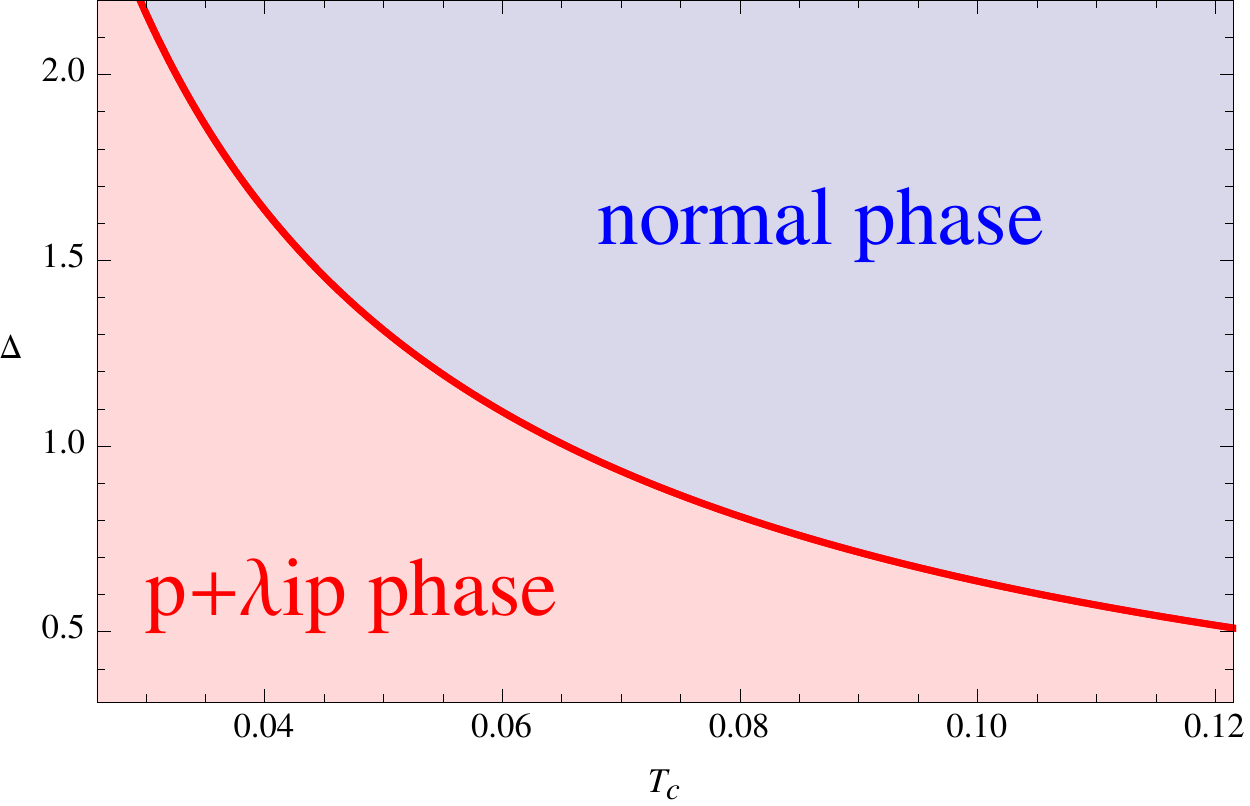}
\caption{\label{DeltaTcPD}$\Delta-T_c$ phase diagram in the probe limit. The bottom left region is occupied by the degenerate p+$\lambda i$p phase and the top right region is occupied by the normal phase.}
\end{figure}

Note that there is an arbitrary valued parameter $\theta$ in the p+$\lambda i$p phases and the result for free energy doesn't depend on it. Therefore at fixed values of $\{\Delta,T\}$, there are infinite number of phases with different values of $\theta$ and they are equally stable. Therefore, the phase diagram of this system in probe limit can be shown as in Figure~\ref{DeltaTcPD}, where we can see that below some critical temperature, the system can be in a p+$\lambda i$p phase with arbitrary value of the angle $\theta$. This means that the superfluid phases are degenerate.

Recall that in Ref.~\cite{Gubser:2008wv}, the p+$i$p phase is unstable. This can be explained by the additional non-abelian coupling between the p and $i$p orders in that model. That nonlinear term will have a positive contribution to the free energy. Therefore the p+$i$p phase in Ref.~\cite{Gubser:2008wv} becomes unstable compared to the p-wave phase. In our setup, the p and $i$p orders are not coupled directly in the action, therefore, all the p+$\lambda i$p phases are all equally stable in the probe limit.

\section{Splitting the degenerate phases with back-reaction}\label{sect:backer}
While we have studied the p+$\lambda i$p phases in the probe limit and find a degenerate behavior, it will be quite interesting to go away from the probe limit and see what will happen to the degenerate phases under the influence of the gravitational interactions between the matter fields. It is easy to notice that when we consider the full interactions between the matter and gravitational fields, the metric will no longer be isotropic in $x-y$ plane in general p+$\lambda i$p phases, except the p+$i$p phase with $\Psi_x=\Psi_y$.

When the back-reaction of matter fields is considered, we can easily get the p-wave solution with $\Psi_y=0$ according to the results in previous study. In addition, we can also find the 
p+$i$p solution because the condition $\Psi_x=\Psi_y$ further implies $f(r)=1$, and therefore the remaining equations of motion are similar to those of the s-wave holographic model. By solving the equations of motion with proper boundary conditions, we can find these two types of solutions, and we draw the condensate behavior in some typical cases.

In the p-wave phase, the condensate value can be extracted by the expectation value $<\mathcal{O}>=\Psi_{x+}$. In order to compare the condensate value of the p+$i$p phase with that of the p-wave phase, we need to define the order parameter in the p+$i$p phase in a consistent way. According to the experience in the probe limit, we can define this value by
\begin{eqnarray}
<\mathcal{O}>=\sqrt{\Psi_{x+}^2+\Psi_{y+}^2}.
\end{eqnarray}
By this definition, the condensate values for the p+$\lambda i$p phases in the probe limit are all equal.

\subsection{Different condensate behaviors}
We can draw the curve of condensation value $<\mathcal{O}>^{1/\Delta}$ versus the temperature $T/\mu$ for various cases. In this section, we show some typical condensate behavior of the p-wave and p+$i$p phases and make comparison.

\begin{figure*}\center
\includegraphics[height=2.8cm] {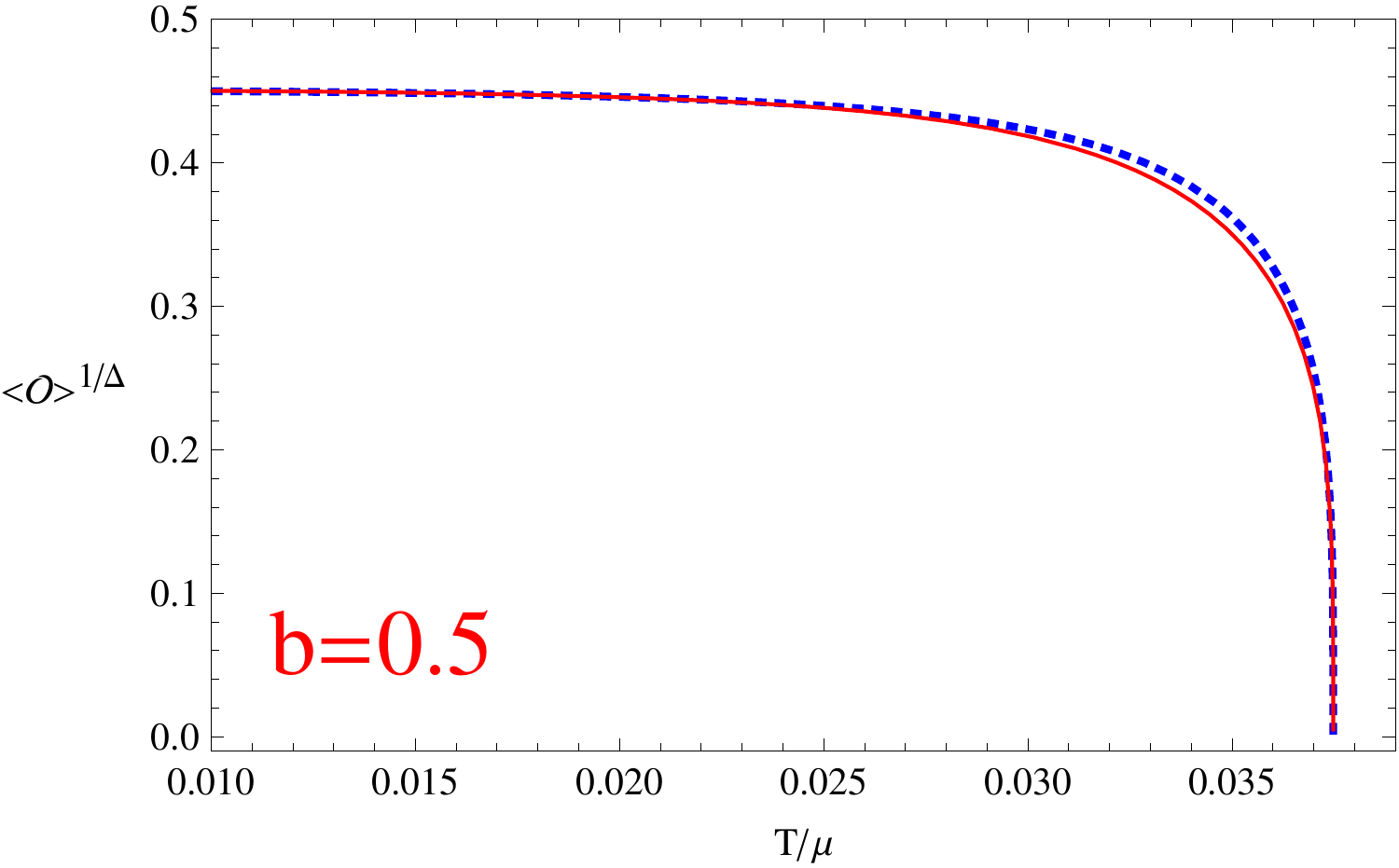}
\includegraphics[height=2.8cm] {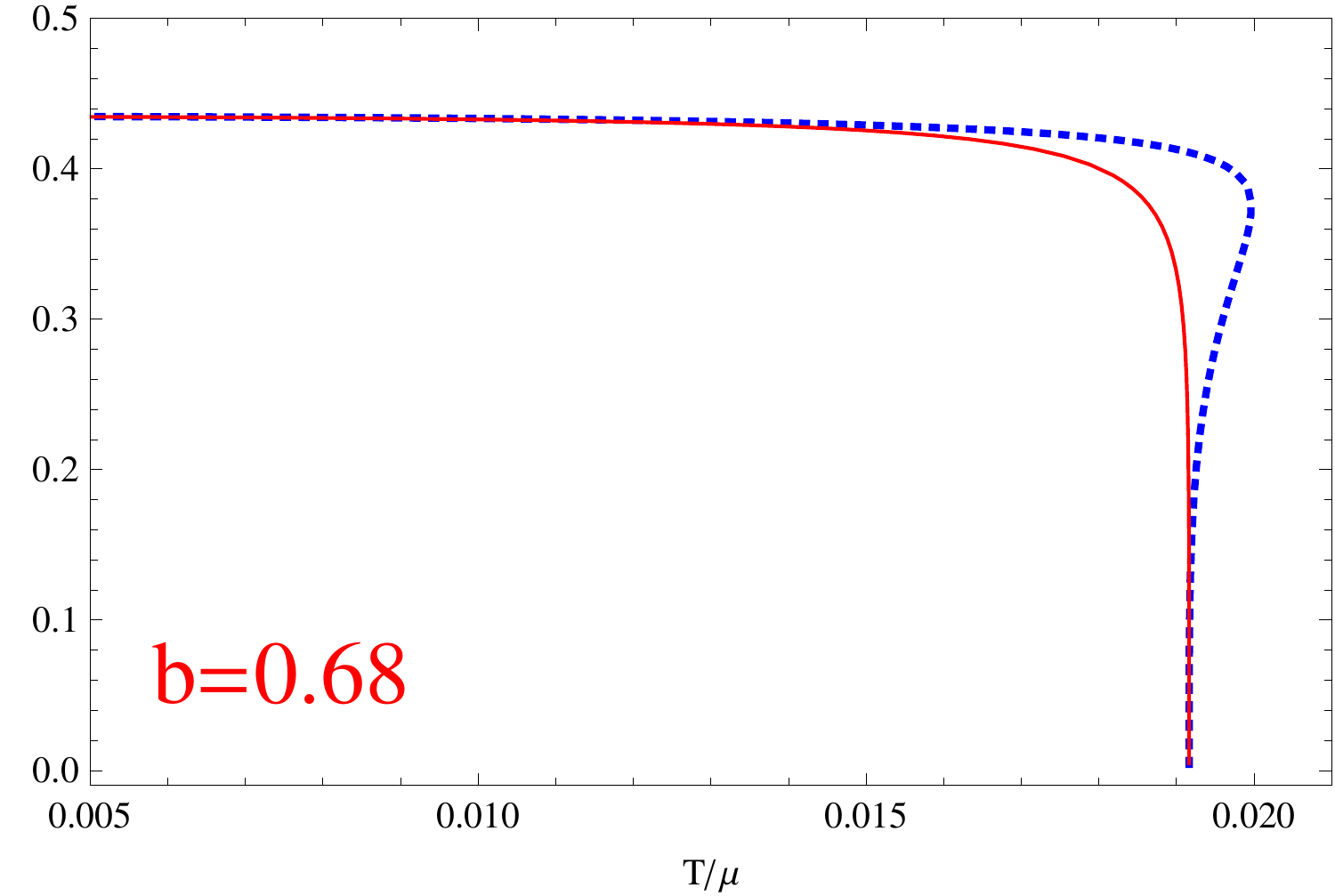}
\includegraphics[height=2.8cm] {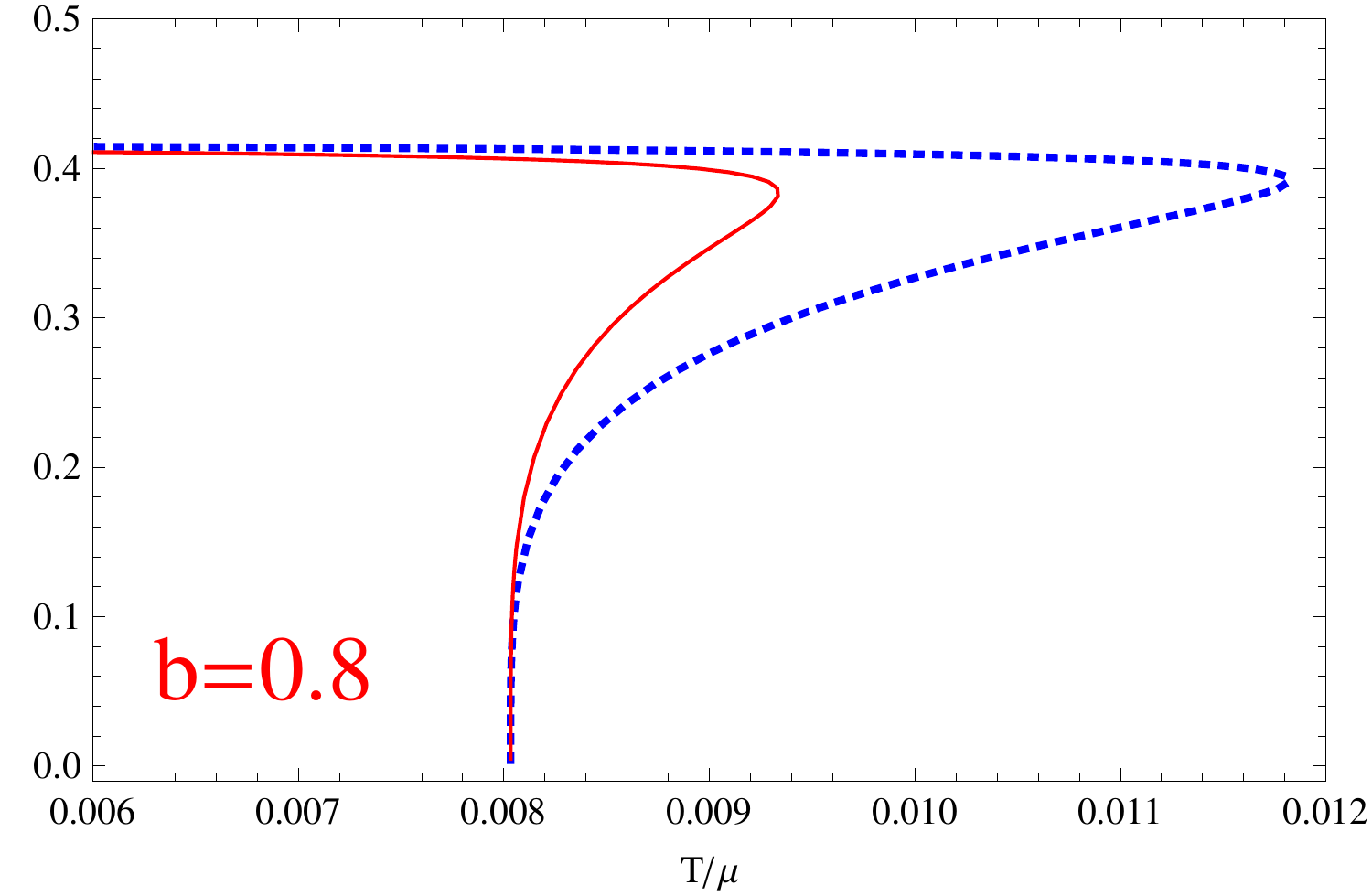}
\caption{\label{CCm0}Condensate behavior with $m_p^2=0$. The dotted blue curves denote the condensate of the p-wave solution and the solid red curves denote the condensate of the p+$i$p solution.}
\end{figure*}

In Figure~\ref{CCm0}, we draw the condensate of the p-wave and p+$i$p solutions with $m_p=0$ and three different values of back reaction strength $b=0.5,~0.68,~0.8$ respectively. The figure with $b=0.5$ shows the typical behavior in the weakly back reacted case where both the p-wave and p+$i$p phase transitions are second order. We can see that the condensate values of the p-wave and p+$i$p phases are very close to each other, and the condensate value of p-wave phase is always larger than that of the p+$i$p phase. In the case that the back reaction is weaker, the difference between the two condensate values is smaller. This is consistent with the results in the probe limit, where the condensate values of the p-wave and p+$i$p phases are equal.

From previous study, we know that the p-wave phase transition will become first order when $b>0.62$. It is interesting to study whether the p+$i$p phase transition will also become first order. Our study show that with $m_p=0$, the p+$i$p phase transition becomes first order when $b>0.69$. Then in the cases $0.62<b<0.69$, the p-wave phase transition becomes first order, while the p+$i$p phase transition is still second order. We show the condensate behavior of such a case with $b=0.68$ in the second plot in Figure~\ref{CCm0}. And in the third plot, we show the case where both the p-wave and p+$i$p phase transitions are first order with $b=0.8$.

\begin{figure*}\center
\includegraphics[height=2.8cm] {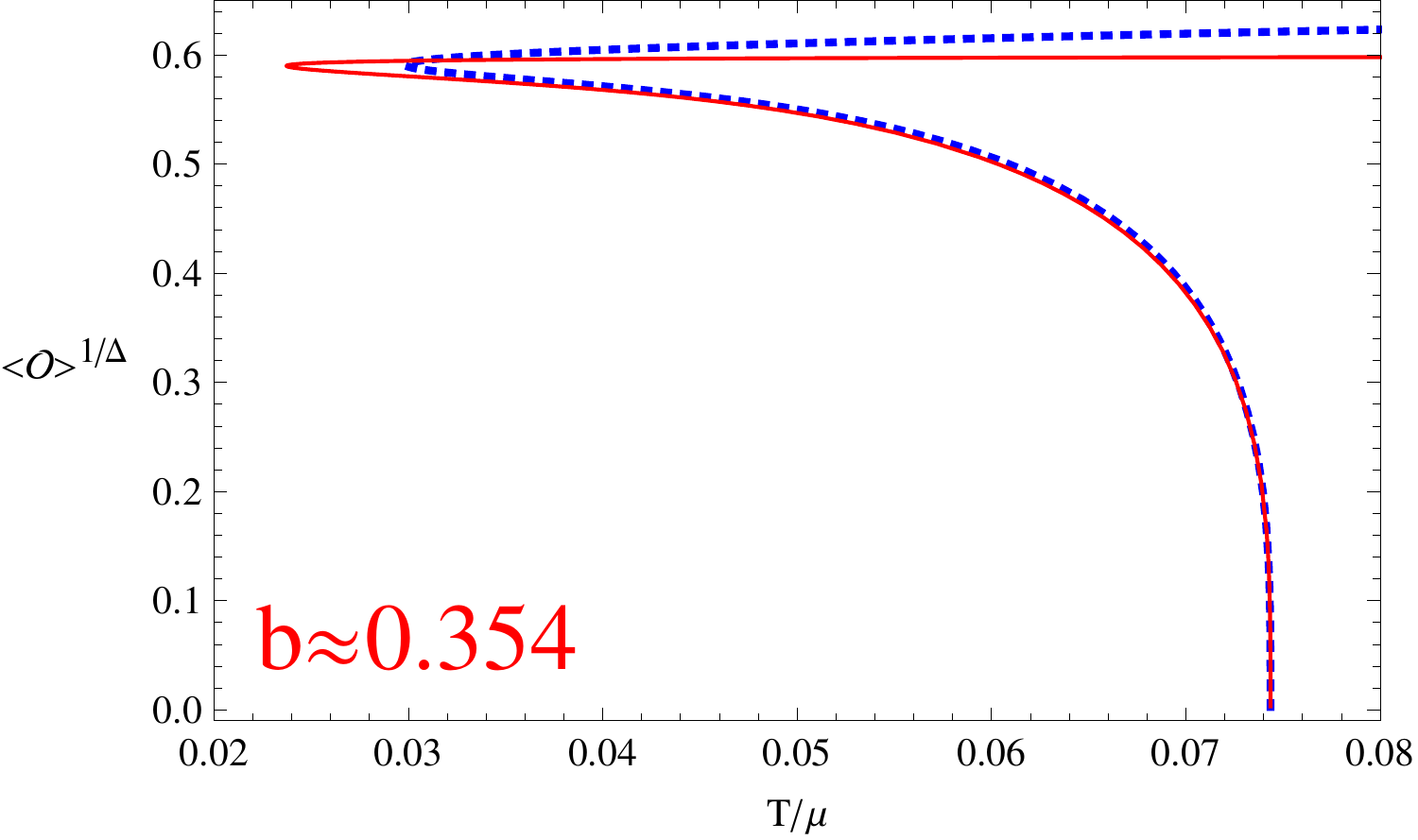}
\includegraphics[height=2.8cm] {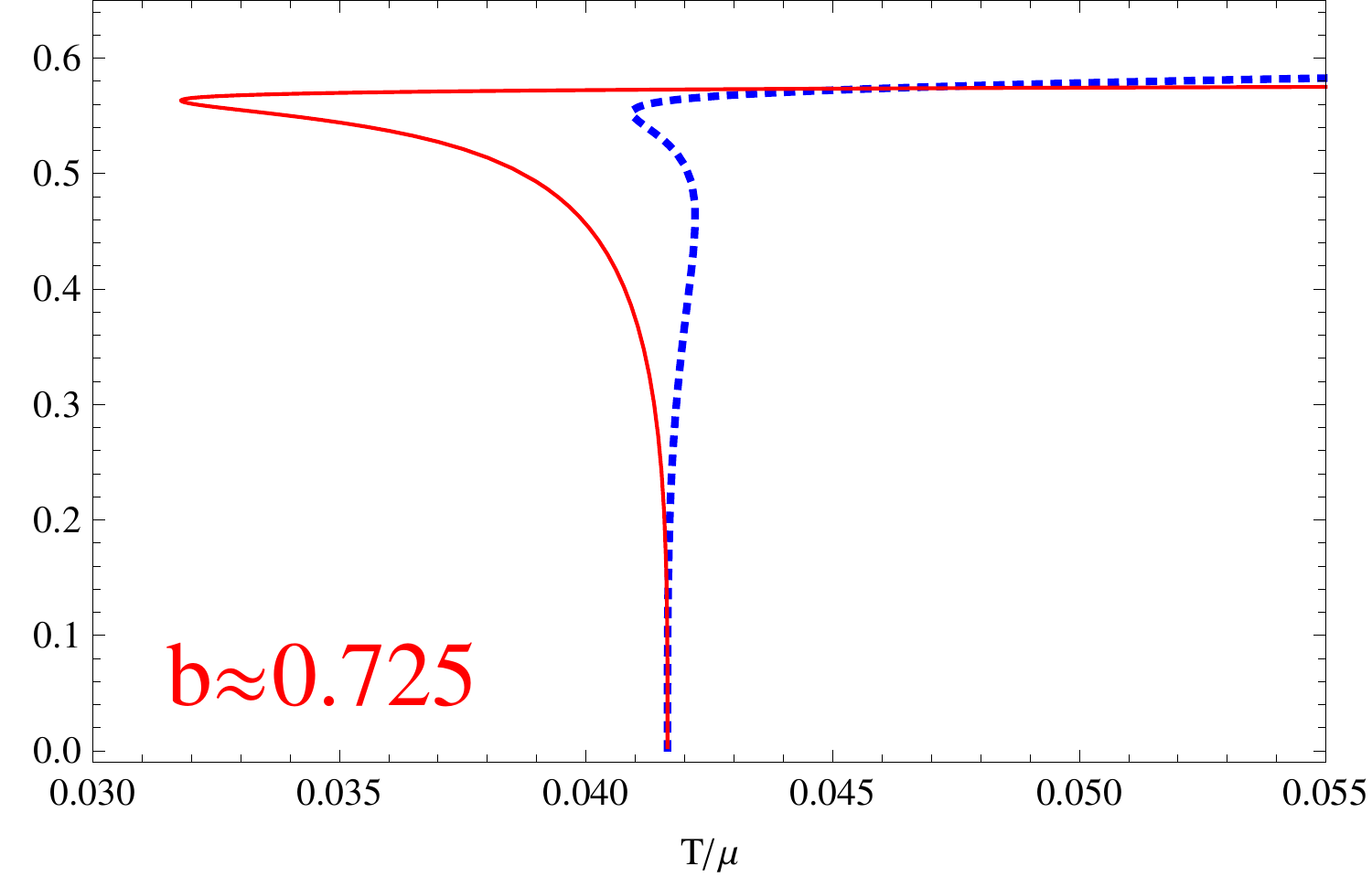}
\includegraphics[height=2.8cm] {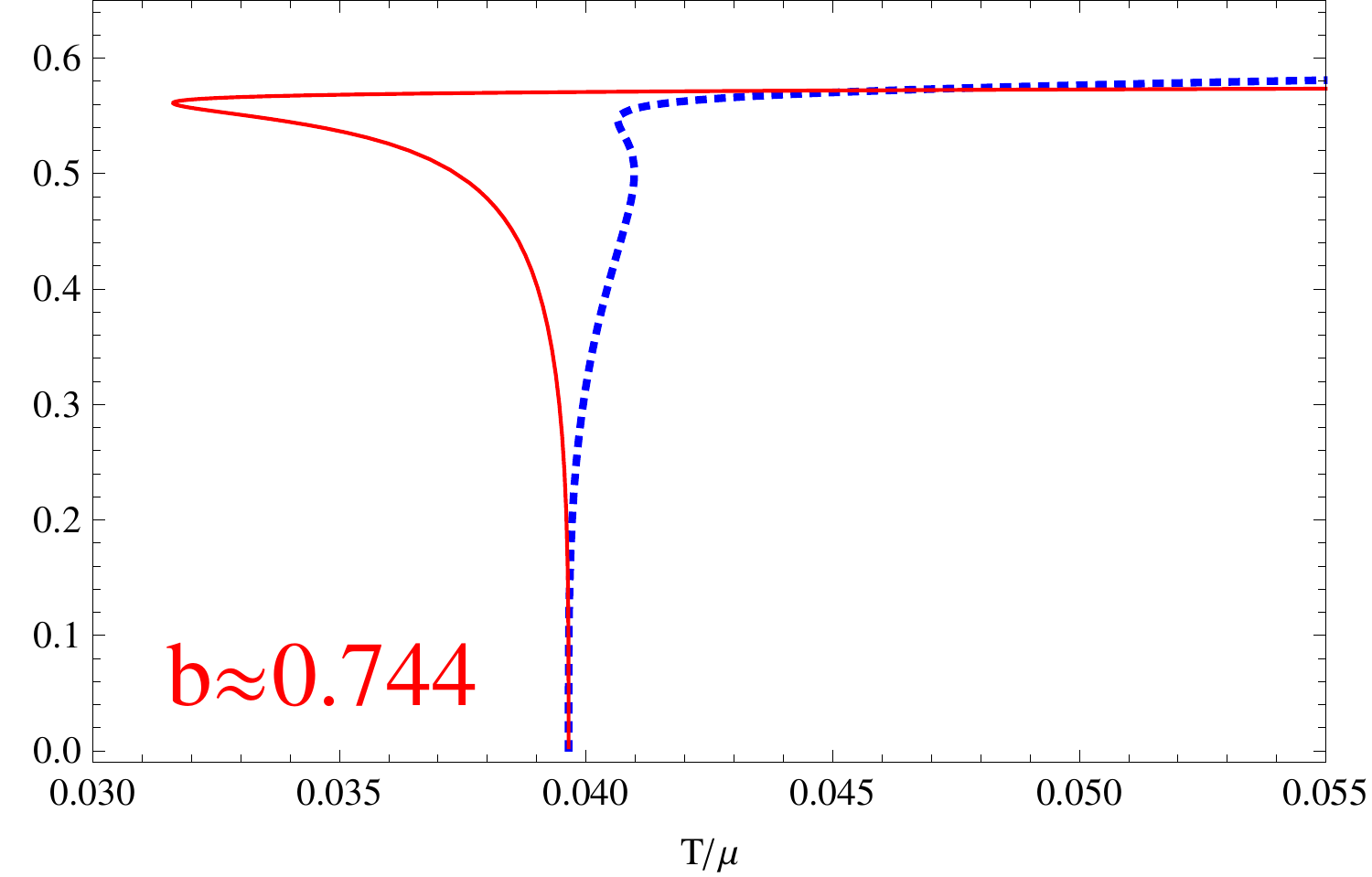}
\caption{\label{CCm2m01875}Condensate behavior with $m_p^2=-3/16$. The dotted blue curves denote the condensate of the p-wave solution and the solid red curves denote the condensate of the p+$i$p solution.}
\end{figure*}

In Ref.~\cite{Cai:2013aca}, the authors showed that the p-wave phase transition will show interesting behavior such as the zeroth order phase transition with $m_p^2=-3/16$. It is also interesting to see what will happen to the p+$i$p phase in this cases. We show our results with $m_p^2=-3/16$ in Figure~\ref{CCm2m01875} with three different values of back reaction parameter $b\approx 0.354, 0.725, 0.744$ respectively. We can see that similar to the p-wave phase transition, the condensate behavior for the p+$i$p phase also signals the zeroth order phase transition back to the normal phase at a lower temperature. But to verify this zeroth order phase transition, we need to calculate the free energy of the p+$i$p solution in these cases. We will show these results in the next subsection.

From all these condensate behaviors of the two phases, we can see a common feature, which is that the two different solutions share the same temperature where the condensate begins to emerge. This can be understood as follows. When the condensate begins to emerge, the amplitude of the condensed field is sufficiently small, therefore we can take this as another kind of probe limit where the fields $\Psi_x$ and $\Psi_y$ are too small to change the metric of a RN-AdS black brane. Then similar to the discussion in the previous section, the symmetry on the $x-y$ plane makes the p+$i$p solution start to emerge at the same temperature as that for the p-wave phase. But when the condensate value becomes large, the metric in the p-wave solution is deformed to be anisotropic while the metric in the p+$i$p solution is still isotropic, this difference results in the difference in condensate as well as the free energy at last.

\subsection{Free energy and the split phases}
To further verify the zeroth order phase transition, and find out which one between the two condensed solutions is the most stable, we need to calculate their free energy. To calculate the free energy holographically, we again calculate the on-shell Euclidean action of the system. But in this section, we should also calculate the contribution of gravity fields. We also need to add some boundary terms such as the Gibbens-Hawking term and boundary counter terms. Because we work in the grand canonical ensemble, the final formula for grand potential can be expressed as
\begin{eqnarray}
\Omega=T S_{E}&=&-\frac{1}{2 \kappa_g ^2}\int d^{4}x \sqrt{-g} (R-2\Lambda)  \\  \nonumber
&&-\frac{1}{\kappa_g ^2}\int d^{3}x \sqrt{-h} (K-\frac{2}{L}),
\end{eqnarray}
where $h$ is the determinant of the induce metric $h_{\mu\nu}$ on the boundary $r\rightarrow \infty$, and $K$ is the trace of the extrinsic curvature $K_{\mu\nu}$.

Using the equations of motion, we can reduce the bulk integrand to total derivative terms of $r$, therefore the final formula only contains boundary terms. The Euclidean time integral will contribute to a factor of $1/T$ and we define the transverse integral on the $x$ and $y$ directions to be the boundary volume $V_2=\int dx dy$. Then the grand potential can be written as
\begin{eqnarray}
\frac{\kappa_g^2 \Omega}{V_2}&=- \lim\limits_{r\to\infty} \Big[ & r N(r) \sigma(r) +\frac{r^2 N'(r) \sigma(r)}{2}  \\ \nonumber
&&+ r^2 N(r) \sigma'(r) - 2 r^2 \sqrt{N(r)} \sigma(r) \Big].
\end{eqnarray}

We can further substitute the boundary expansion into the formula to get
\begin{eqnarray}
\frac{\kappa_g^2 \Omega}{V_2}&=-M_{b0}.
\end{eqnarray}

\begin{figure*}\center
\includegraphics[height=2.8cm] {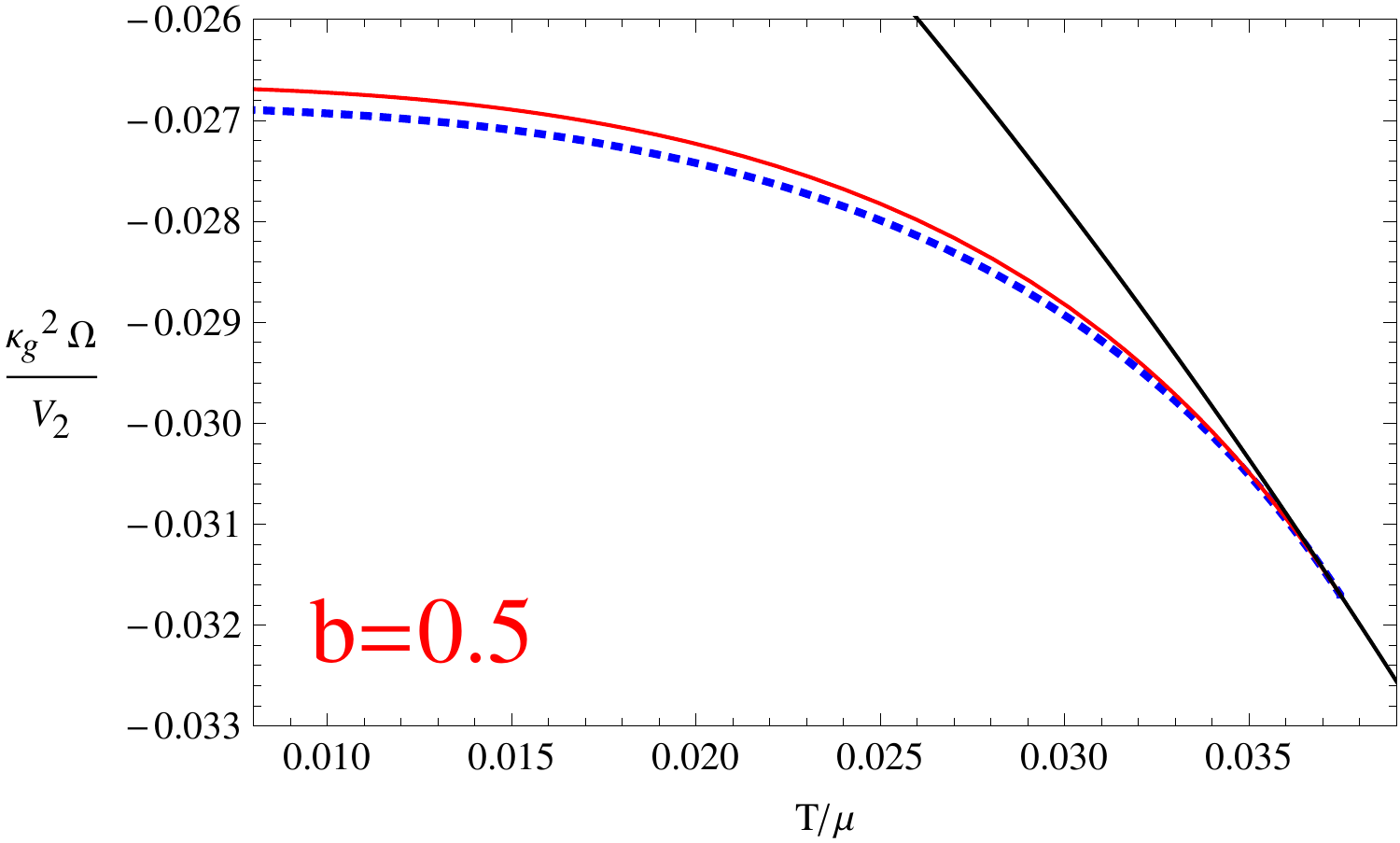}
\includegraphics[height=2.8cm] {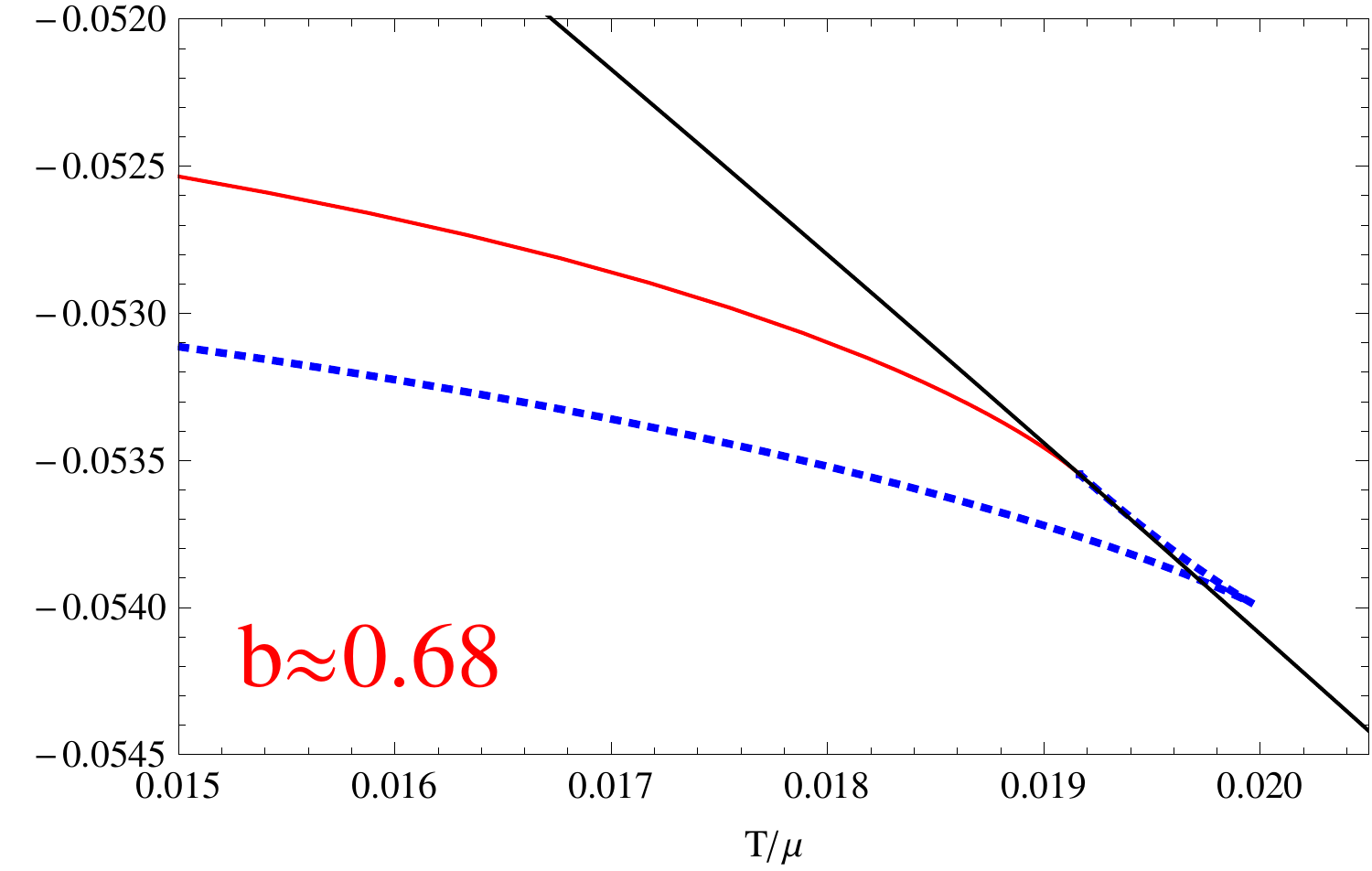}
\includegraphics[height=2.8cm] {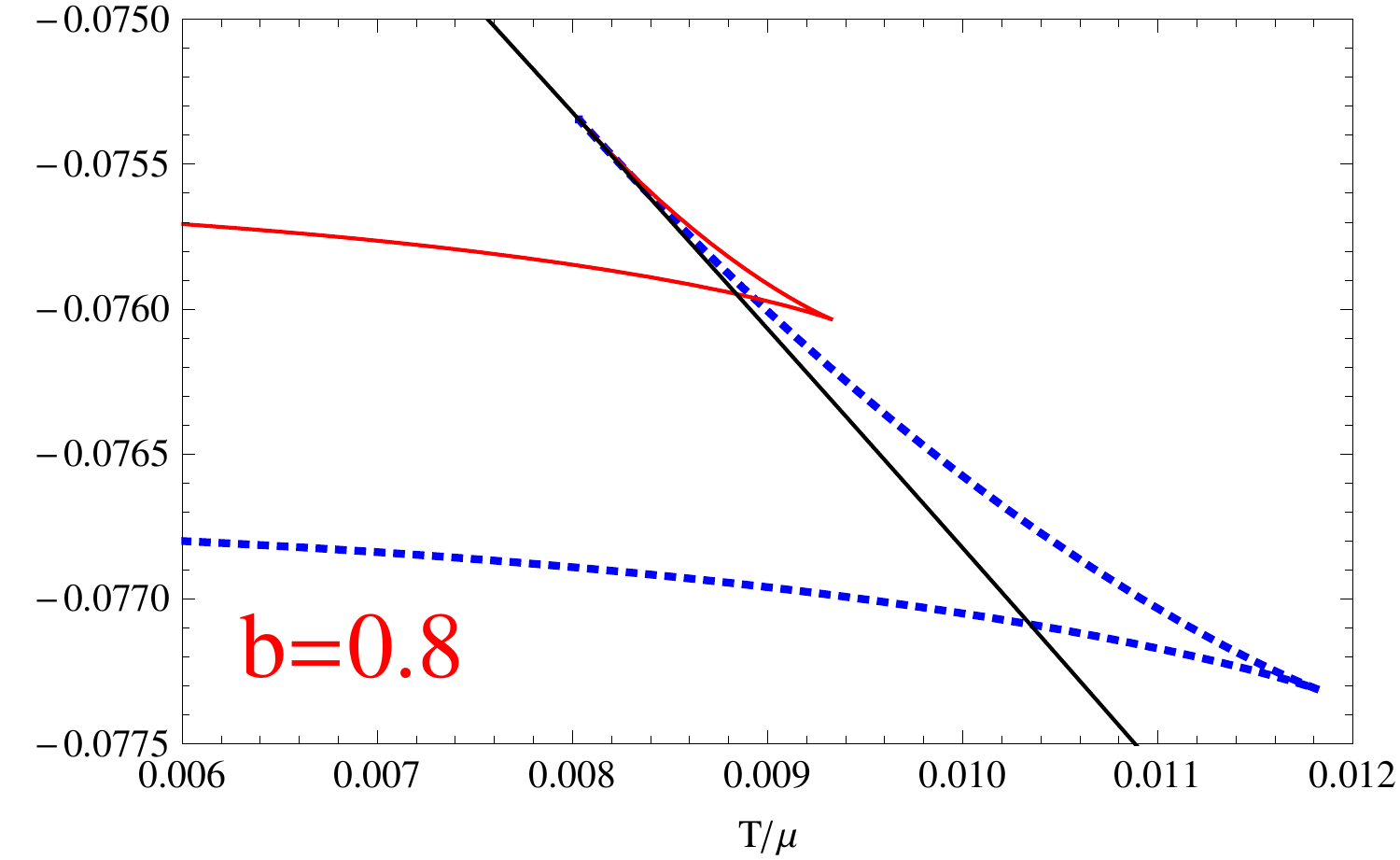}
\caption{\label{FreeEm0}Free Energy curves with $m_p^2=0$. The dotted blue curves denote the free energy of the p-wave solution, the solid red curves denote the free energy of the p+$i$p solution and the solid black curves denote the free energy of the normal phase.}
\end{figure*}

\begin{figure*}\center
\includegraphics[height=2.8cm] {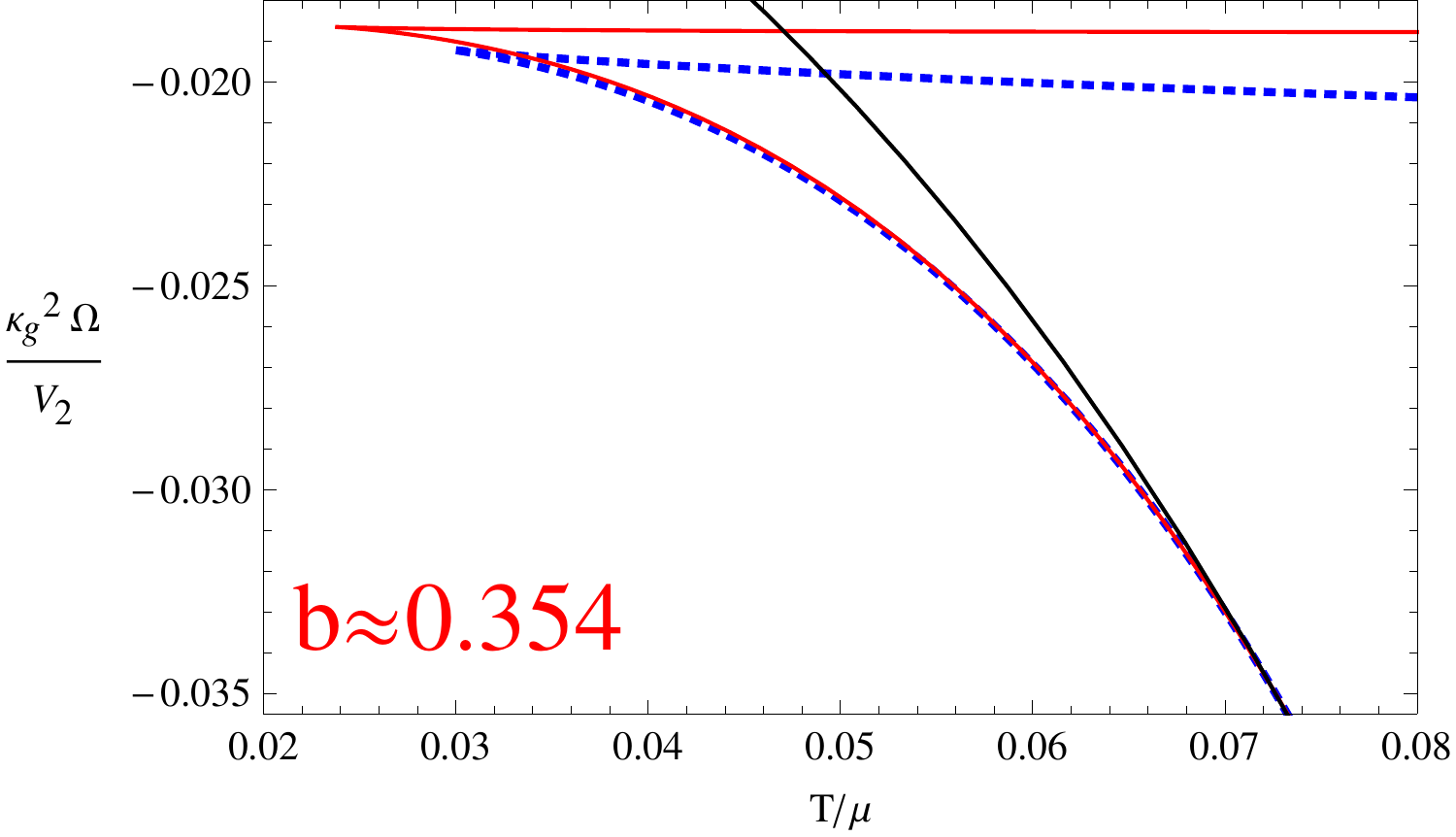}
\includegraphics[height=2.8cm] {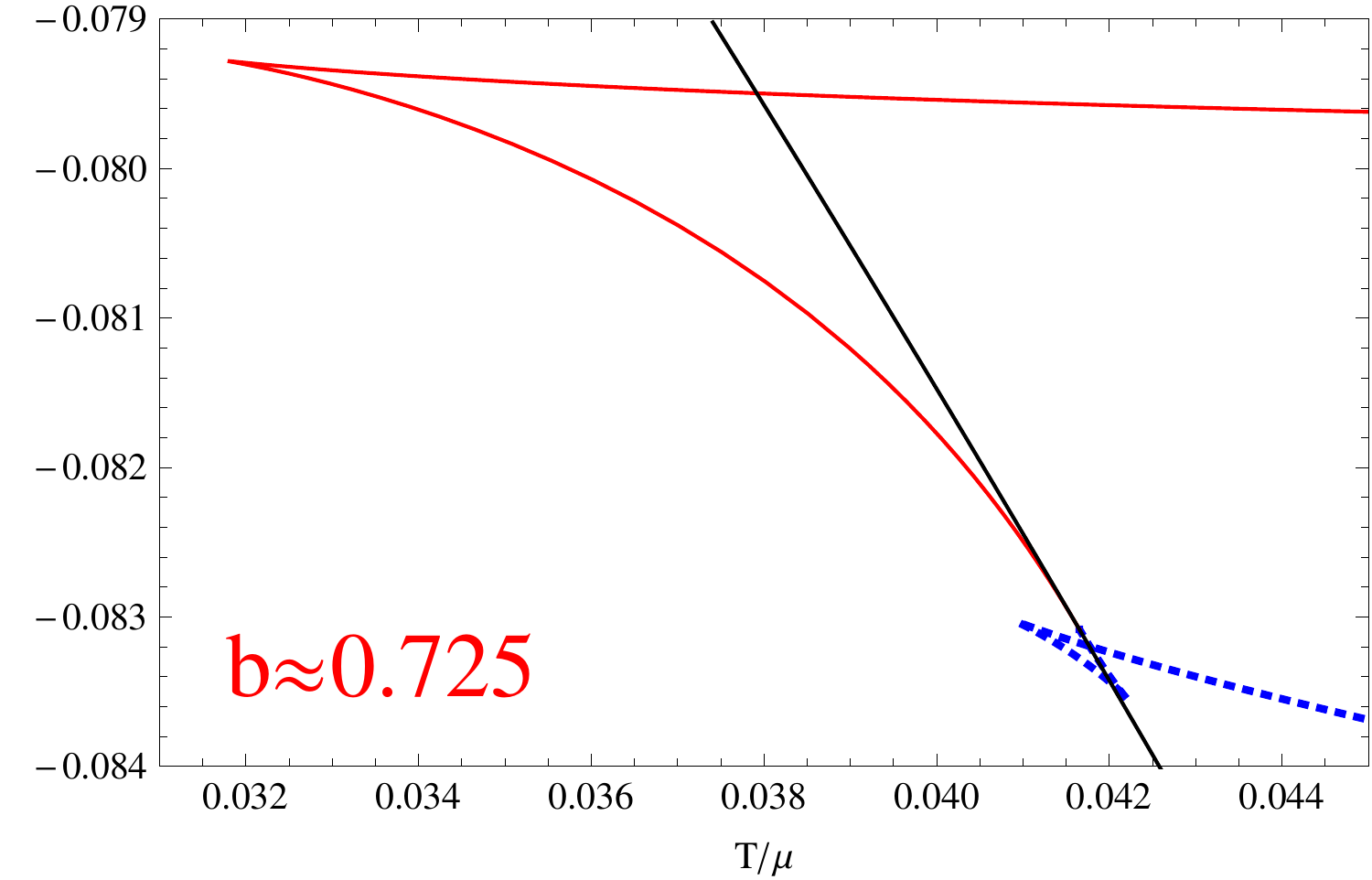}
\includegraphics[height=2.8cm] {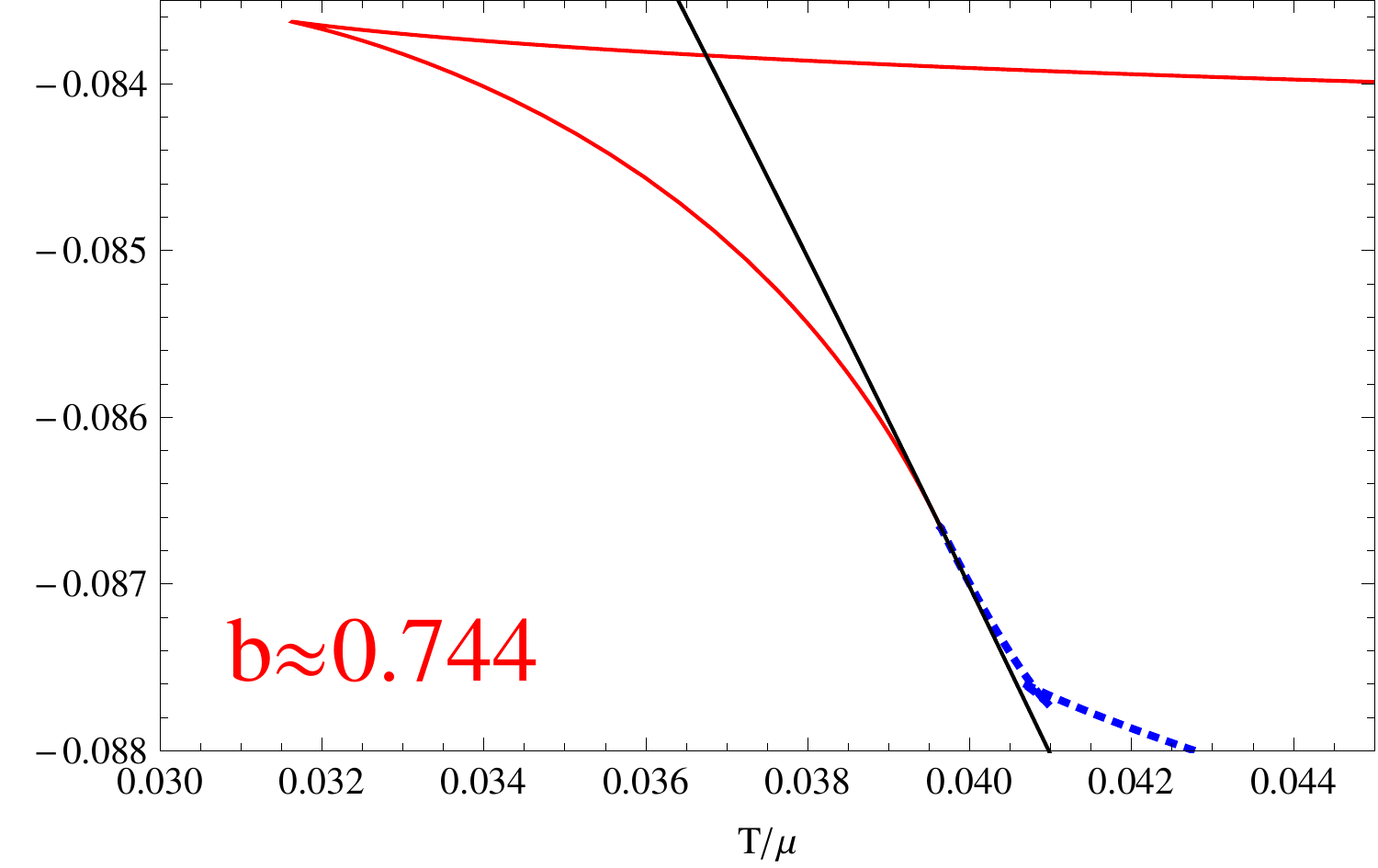}
\caption{\label{FreeEm2m01875}Free Energy curves with $m_p^2=-3/16$. The dotted blue curves denote the free energy of the p-wave solution, the solid red curves denote the free energy of the p+$i$p solution and the solid black curves denote the free energy of the normal phase.}
\end{figure*}

We draw the curves of the free energy in Figure~\ref{FreeEm0} and Figure~\ref{FreeEm2m01875}. In Figure~\ref{FreeEm0}, we show the free energy curves for the cases with $m_p^2=0$. We can see that in these cases, no matter the phase transitions are first order or second order, the free energy for the p-wave phase is always lower than that of the p+$i$p solution at the same temperature. This indicates that with $m_p^2=0$, although we can find p+$i$p solutions, these solutions are always unstable.

In Figure~\ref{FreeEm2m01875}, we show the free energy curves for the cases with $m_p^2=-3/16$. From these three plots, we can confirm that the free energy changes non-continuously at the left side of the condensed phases, and the phase transition back to the normal phase is zeroth order. We can also compare the free energy of the two condensed phases. We can see that in the temperature region where both the p-wave and p+$i$p phases exist, the free energy of the p-wave phase is still lower. However, in these cases, there is some temperature region where only the p+$i$p phase exists, and in this region, the p+$i$p phase has the lowest free energy. Therefore, we can claim that we succeed on building stable holographic p+$i$p phase with considering the back-reaction on metric.

We can also analyze more details of the cases with $m_p^2=-3/16$. From the plot with $b\approx 0.354$ we can see that if we lower the temperature of the system from a high value, the system will undergo three phase transitions. The first one is the second order phase transition form the normal phase to the p-wave phase, and the second one is a zeroth order phase transition from the p-wave phase to the p+$i$p phase. The final one is a zeroth order phase transition from the p+$i$p phase back to the normal phase. In the case with $b\approx 0.725$ the situation is similar, except that the first phase transition from normal phase to the p-wave phase becomes first order. In last plot with $b\approx 0.744$, the p-wave phase is always unstable compared to the normal phase, therefore the system will undergo only two phase transitions. The first one is from the normal phase to the p+$i$p phase and the second is the zeroth order phase transition from p+$i$p phase to the normal phase.

\section{\bf Conclusions and discussions}\label{sect:conclusion}
In this paper, we setup the p+$i$p phase in the holographic p-wave model with complex vector fields in the bulk. We found that in the probe limit, due to the symmetry between the x and y directions, we can construct a set of p+$\lambda i$p phases, and the results of free energy show that these phases with different values of $\lambda$ are equally stable.

We also extend the study to the cases away from probe limit, and found that with considering the back reaction on the metric fields, the p-wave and p+$i$p solutions get different values of free energy. In all the temperature region where the two solutions both exist, the p-wave one gets a lower value of free energy. It seems that the effect of back reaction is to split the degenerate p+$\lambda i$p phases to a p-wave solution with lower free energy and a p+$i$p solution with higher free energy.

It is also very interesting that because of the zeroth order phase transition from the p-wave phase has a higher critical temperature than that from the p+$i$p solution, the p+$i$p solution exists in a temperature region where no p-wave phase exists. In this region, the free energy of p+$i$p phase is still lower than that of the normal phase, thus the p+$i$p phase can be stable. This is the first time that one builds a stable p+$i$p superfluid phase holographically. 

There are also many further extensions from our results that are very interesting. For example, we are going to consider this topic in Einstein-Gauss-Bonnet gravity, and see whether the stable p+$i$p phase can be found where the p-wave solution exists. We can also consider the competition between the p+$i$p phase and the s-wave phase to compare the two isotropic solutions. Although the p+$i$p phase is isotropic in 2+1 dimensions, it becomes anisotropic in 3+1 dimensions. It would be interesting to study the anisotropic behaviors such as conductivity of the p+$i$p phase and compare the results with the p-wave case from a 4+1 dimensional bulk. Finally, one can try to realize more complex p-wave orders holographically, and engineer phase diagrams of a Helium like system. We are looking forward to more progress in these directions.

\begin{acknowledgement}
\section*{Acknowledgements}
ZYN would like to thank Professor Jan Zaanen for useful discussions and suggestions. ZYN and HZ would like to thank the organizers of the program "BLACK HOLES AND EMERGENT SPACETIME" held in Nordic Institute for Theoretical Physics(NORDITA, Stockholm, Sweden) for their hospitality. This work was supported in part by the National Natural Science Foundation of China under Grant Nos. 11565017, 11675140, 11247017, 11275066, 11491240167, 11447131, and 11565016, in part by the Open Project Program of Key Laboratory of Theoretical Physics, Institute of Theoretical Physics, Chinese Academy of Sciences (No. Y5KF161CJ1), in part by a talent-development fund from Kunming University of Science and Technology under Grant No. KKSY201307037, and in part by Hunan Provincial Natural Science Foundation of China under Grant No. 2016JJ1012.
\end{acknowledgement}


\end{document}